\documentclass[aps,prl,reprint,twocolumn,superscriptaddress,floatfix,showpacs,footnoteinbib]{revtex4-1}

\usepackage[utf8]{inputenc}
\usepackage{amsmath}
\usepackage{graphicx,color}
\usepackage{mathptmx}
\usepackage{helvet}
\usepackage{color}

\newcommand{\Arkady}[1]{\textcolor{black}{#1}}

\begin{document}
\title{\Large Radiation-dominated particle and plasma dynamics}

\author{Arkady Gonoskov} 
\affiliation{Department of Physics, Chalmers University of Technology, SE--412 96 G\"oteborg, Sweden}
\affiliation{Institute of Applied Physics, Russian Academy of Sciences, Nizhny Novgorod 603950, Russia}
\affiliation{Lobachevsky State University of Nizhni Novgorod, Nizhny Novgorod 603950, Russia}

\author{Mattias Marklund} 
\affiliation{Department of Physics, Chalmers University of Technology, SE--412 96 G\"oteborg, Sweden}

\begin{abstract}
We consider the general problem of charged particle motion in a strong electromagnetic field of arbitrary configuration and find a universal behaviour: for sufficiently high field strengths, the radiation losses lead to a general tendency of the charge to move along the direction that locally yields zero lateral acceleration. The relativistic motion along such a direction results in no radiation losses, according to both classical and quantum descriptions of radiation reaction. We show that such a radiation-free direction (RFD) exists at each point of an arbitrary electromagnetic field, while the time-scale of approaching this direction decreases with the increase of field strength. Thus, in the case of a sufficiently strong electromagnetic field, at each point of space, the charges mainly move and form currents along local RFD, while the deviation of their motion from RFD can be calculated in order to account for their incoherent emission. This forms a general description of particle, and therefore plasma, dynamics in strong electromagnetic fields, the latter can be generated by state-of-the-art lasers or in astrophysical environments.
\end{abstract}


\maketitle


The development of state-of-the-art high-intensity laser systems has spurred renewed interest in radiation reaction and its effect on particle dynamics in strong electromagnetic fields (see, e.g., Ref.\ \onlinecite{Turcu-etal} and references therein). The nature of radiation reaction has been a long-standing issue in the literature \cite{Rohrlich, Ritus, Baier-book}, and the developments over the last decade \cite{DiPiazza-review,Greger-Anton1,Greger-Anton2, gonoskov.pre.2015, Dinu-etal} have further clarified various aspects of this effect. Furthermore, particle and plasma dynamics in the presence of significant radiation losses appears as a difficult, but rapidly developing, field for theoretical studies \cite{Baier-etal2,DiPiazza-review,Blackburn-etal, DiPiazza, fedotov.pra.2014, esirkepov.pla.2015, kirk.ppcf.2016, s.bulanov.jpp.2017, bogdanova.2017, esirkepov.pla.2017}. The interest to this field has been further increased by the discovery of several somewhat counter-intuitive phenomena, such as straggling \cite{Shen-White,Duclous-etal}, quenching \cite{harvey.prl.2017}, radiation reaction trapping in traveling waves \cite{pukhov.prl.2014}, as well as normal \cite{Kirk:2009vk} and anomalous \cite{gonoskov.prl.2014} radiative trapping in standing waves. Despite continuous efforts on developing analytical approaches \cite{kirk.ppcf.2016, esirkepov.pla.2017, Nerush-etal}, the high degree of nonlinearity in many cases restricts the analysis to qualitative explanations, on the level of cause-and-effect relations, supported by numerical simulations. Therefore general theoretical approaches are needed for building up a more comprehensive picture, useful for developing experimental concepts at upcoming laser facilities \cite{eli-np, eli-beams, XCELS, VULCAN}, where these, and other phenomena due to radiation reaction, can be exploited for creating exotic sources of particles and radiation \cite{ridgers.pop.2013, jirka.pre.2016, grismayer.pop.2016, gonoskov.prx.2017, stark.prl.2016}, as well as extreme states of matter \cite{muraviev.jetp.2015, efimenko.arxiv.2017}.

In this paper, we consider particle dynamics in a strong electromagnetic field. We show that for an arbitrary electromagnetic field, there always exists a direction that yields zero transverse acceleration and thus zero value for the dominant term of the relativistic radiation reaction force. We demonstrate that radiation reaction affects the dynamics of charges so that they tend to move along this direction. If this happens on a timescale comparable to, or shorter than, the timescale of the field variations, particle and plasma dynamics can be self-consistently described through the assumption that the particles follow this radiation-free direction at each point of space.

\textit{The radiation-free direction.\textemdash}Let us consider an arbitrary electromagnetic field and its effect on a charged particle at an arbitrary point of space and time. One may ask if there exists a direction of propagation $\textbf{n}$ for which the particle does not experience any lateral acceleration, i.e. ${\textbf{E}-\left(\textbf{E}\cdot\textbf{n}\right)\textbf{n}+\left(v/c\right)\left(\textbf{n}\times\textbf{B}\right)=0}$, where $v$ and $c$ are the particle's velocity and the speed of light, $\textbf{E}$ and $\textbf{B}$ are the local electric and magnetic field vectors. If a particle move along such a direction with relativistic speed, the particle predominantly does not experience any radiation losses, since the lateral acceleration provides the dominant loss mechanism according to both classical and quantum description. We thus call this the \textit{radiation-free direction} (RFD) and consider the problem of finding such a direction in generic case.

\begin{figure}
	\centering\includegraphics[width=0.5\columnwidth]{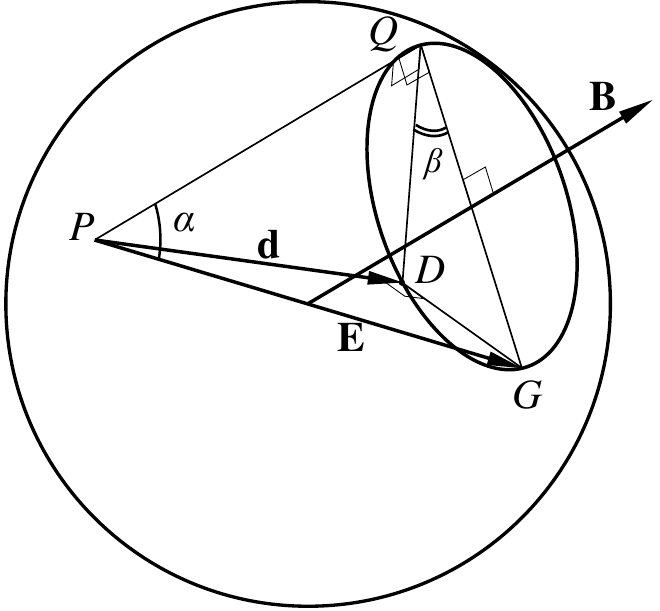}
	\caption{The general setup of the vectors defined in the text.} \label{geometry}
\end{figure}

Suppose $\textbf{P}$ is an arbitrary point in a 3D space. Let us consider a sphere with the centre in the point $\textbf{P} + (1/2)\textbf{E}$ and the radius $\left|\textbf{E}\right|/2$ (see fig.~\ref{geometry}). We associate each point $\textbf{D}$ of this sphere with the direction of propagation orientated along the vector $\textbf{d} = \textbf{D} - \textbf{P}$. We now define a point $\textbf{G} = \textbf{P} + \textbf{E}$. The vector $\textbf{F}_{\bot}^E = \textbf{D} - \textbf{G}$ is then the transverse component, relative to the particle direction of motion, of the Lorentz force due to the electric field (here we for simplicity assume the charge to be equal to unity; this can easily be reinstated when necessary). 

By definition, the magnetic part of the Lorentz force $\textbf{F}^B$ is always perpendicular to both the magnetic field vector $\textbf{B}$ and the direction of the particle motion. Thus the transverse component $\textbf{F}^B_{\bot}$ coincides with $\textbf{F}^B$ and is perpendicular to the vector $\textbf{B}$. In order for $\textbf{F}^B_{\bot}$ to be cancelled by the transverse component of the electric force term $\textbf{F}_{\bot}^E$, the latter has to be perpendicular to the magnetic field vector. This is the reason why we are restricted to consider directions that correspond to point at the circle that is intersection of the sphere with the plane passing through the point $\textbf{G}$ and being perpendicular to the vector $\textbf{B}$. For all points of this circle $\textbf{F}_{\bot}^E {\bot} \, \textbf{B}$.

For all the points at that circle, the vectors $\textbf{F}^B_{\bot}$ and $\textbf{F}^E_{\bot}$ are not only lying in the plane perpendicular to $\textbf{B}$, but also parallel to each other. (The only exception is the case of the vector $\textbf{d}$ being parallel to the vector $\textbf{B}$, which results in $\textbf{F}^B_{\bot} = 0$ and thus does not contradict  the final conclusions of this section.) Indeed, if the vectors $\textbf{d}$ and $\textbf{B}$ are not parallel, then for the planes perpendicular to these vectors and passing through the point $\textbf{D}$ there is a single line of intersection. Both vector $\textbf{F}^B_{\bot}$ and vector $\textbf{F}^E_{\bot}$ have to be parallel to this line as each of them lies in both planes. Thus $\textbf{F}^B_{\bot}$ and $\textbf{F}^E_{\bot}$ are parallel to each other. 

As a result, in order for the particle to have zero lateral acceleration we only need to request the absolute values of $\textbf{F}^B_{\bot}$ and $\textbf{F}^E_{\bot}$ to be equal to each other (as we will see one of the solutions of this problem corresponds to the opposite orientations of the vectors so that they cancel each other). 

Let us consider a point $\textbf{Q}$, which belongs to the circle and is opposite to the point $\textbf{G}$. The segment $\textbf{PQ}$ is parallel to $\textbf{B}$. We define $\alpha$ as the angle between the vectors $\textbf{E}$ and $\textbf{B}$. The length of the $\textbf{QG}$ segment is then $\left|\textbf{QG}\right| = E \sin \alpha$. We define $\beta$ as the angle between segments $\textbf{QG}$ and $\textbf{QD}$. The value of $\beta \in \left[-\pi/2, \pi/2\right]$ defines the position of the point $\textbf{D}$ on the circle. Then we can write $
\left| \textbf{F}^E_{\bot} \right| = \left| \textbf{DG} \right| = E \sin \alpha \left| \sin \beta \right| .$

Suppose $\textbf{v}$ is the particle's velocity vector, and $\textbf{v}_{\bot}$ is its component across the vector $\textbf{B}$. Then the ratio $|\textbf{v}_{\bot}|/|\textbf{v}| = v_{\bot}/v$ is equal to the ratio of the lengths of the segment $\textbf{QD}$ to the one of the segment $\textbf{PD}$ in the right-angled triangle $\textbf{PQD}$. As a result we obtain
\begin{equation}
\left| \textbf{F}^B_{\bot} \right| = \frac{v}{c} \frac{v_{\bot}}{v} B = \frac{v}{c} B \frac{\sin \alpha \cos \beta}{\sqrt{\cos^2 \alpha + \sin^2 \alpha \cos^2 \beta}}.
\end{equation}
Equating $\left|\textbf{F}^B_{\bot}\right|$ and $\left|\textbf{F}^E_{\bot}\right|$, we obtain the equation $
u^2\sin^2\alpha = u \left(k + 1\right) - k $ 
for the unknown quantity $u = \sin^2 \beta$, where $k = (v^2/c^2) (B^2/E^2)$. One can show that in the range $0 < u < 1$ this equation always has only one solution, which is
\begin{equation}
\label{u_def}
u = \frac{\left(\sqrt{k} + 1/\sqrt{k}\right) - \sqrt{\left(\sqrt{k} + 1/\sqrt{k}\right)^2 - 4 \sin^2 \alpha}}{2\sin^2 \alpha/\sqrt{k}}.
\end{equation}
Assuming $v \approx c$ and using $\left(\textbf{E} \times \textbf{B}\right)^2 = E^2 B^2 \sin^2\alpha$, we can express the solution in ultra-relativistic case:
\begin{equation}
u = 2 \frac{B^2}{E^2 + B^2}\frac{1 - \sqrt{1 - w}}{w}, \: \: w = \frac{4 \left(\textbf{E} \times \textbf{B}\right)^2}{\left(E^2 + B^2\right)^2}.
\label{u_rel} 
\end{equation} 

\begin{figure}
\centering\includegraphics[width=0.7\columnwidth]{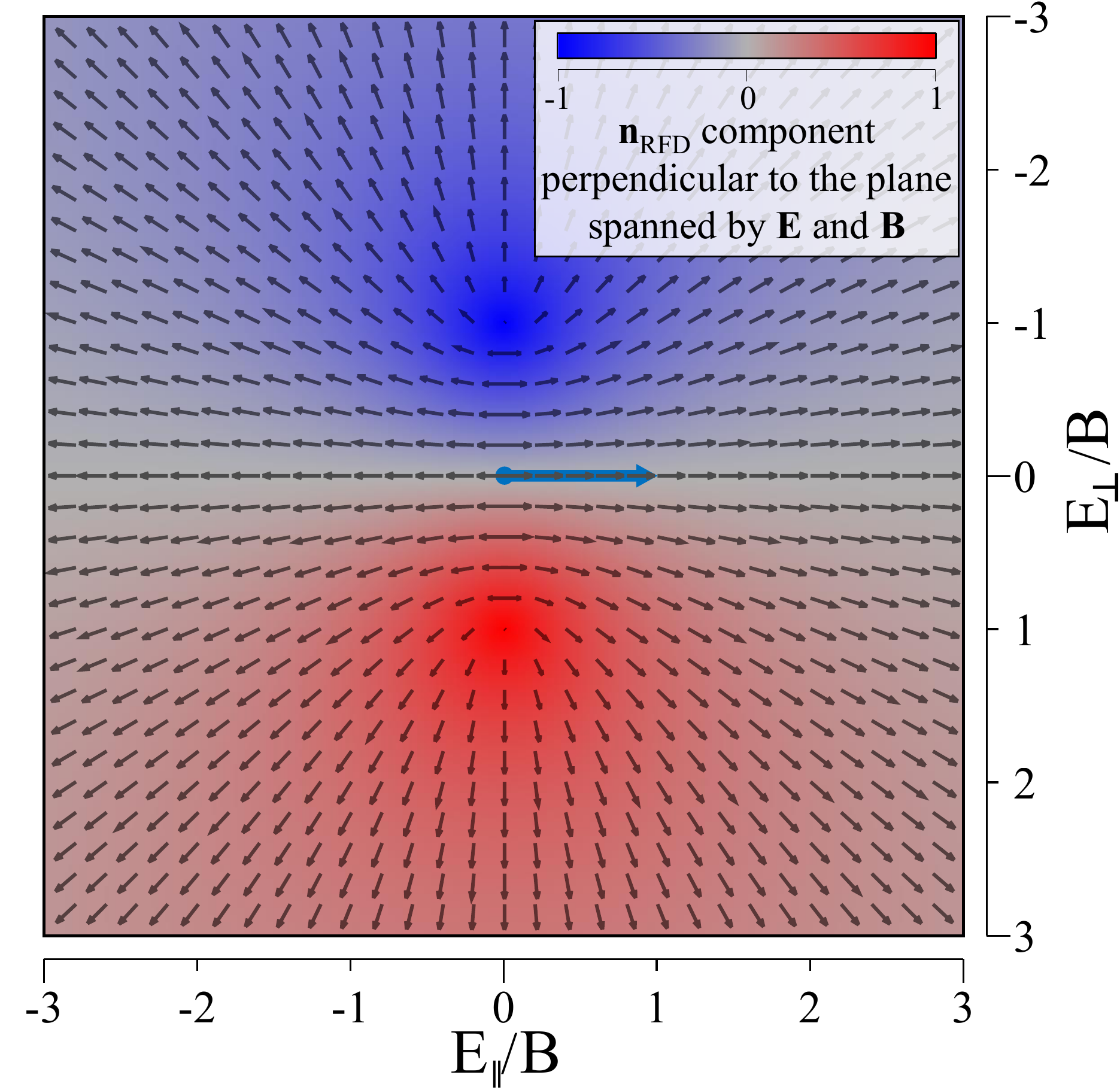}
\caption{The radiation-free direction functional $\textbf{n}_{\rm RFD}^+\left(\textbf{E}, \textbf{B}\right)$ shown by fixing $\textbf{B}$ (blue arrow) and varying the end point of $\textbf{E}$, keeping its starting point in the centre. For each position of the end point the gray arrow shows the projection of $\textbf{n}_{\rm RFD}^+$ on the plane spanned by ${\bf E}$ and ${\bf B}$, while the colour denotes the component perpendicular to this plane. The result for the negative charge ($\textbf{n}_{\rm RFD}^-\left(\textbf{E}, \textbf{B}\right)$) can be obtained through inverting the component perpendicular to $\textbf{E} \times \textbf{B}$.} \label{rfd_view}
\end{figure}

As one can see from the expression for $w$, its maximal value is equal to unity when the electric and magnetic fields are equal and perpendicular to each other. Thus, the solution is always real. For the obtained value of $u$ there are two values of $\beta$ (one negative and one positive) that give $\sin^2\beta = u$. However, exactly one of them corresponds to the case of the opposite orientation of the electric and magnetic components, so that they cancel each other. Using fig.~\ref{geometry}, we can derive an explicit form of the functional $\textbf{n}_{\rm RFD}^{\pm}\left(\textbf{E}, \textbf{B}\right)$ that defines the direction of motion with zero lateral acceleration:
\begin{equation} \label{eq:rdf}
\textbf{n}_{\rm RFD}^{\pm}\left(\textbf{E}, \textbf{B}\right) = \frac{\sqrt{u - u^2} \,B\left(\textbf{E} \times \textbf{B} \right) \pm \left[\left(1 - u\right)B^2\textbf{E} + u \left(\textbf{E} \cdot \textbf{B}\right) \textbf{B}\right]}{B\sqrt{E^2 B^2  - u \left(\textbf{E} \times \textbf{B} \right)^2}},
\end{equation}
where the superscript denotes the sign of the charge and $u$ is defined through Eq.~(\ref{u_def}) in general case or Eqs.~(\ref{u_rel}) in ultra-relativistic case. As one can see the problem has exactly one solution in all cases except the case of ($\textbf{E} \bot \textbf{B}$, $E < B$), when there are two solutions. The functional is illustrated in fig.~\ref{rfd_view}. 

\textit{Approaching the radiation-free direction.\textemdash}Radiation reaction gives a change of the particle's momentum, and this change is always orientated exactly against the direction of propagation. Thus, radiation reaction itself cannot be directly responsible for the change of the direction of motion. However, radiation reaction does reduce the energy of a particle: this is essential, since particles with higher energies resist changing their direction of motion due to their large relativistic inertia. Thus, radiation reaction changes the direction of propagation indirectly, through reducing the particle energy, so that the Lorentz force starts to affect more strongly the direction of propagation. We will now show that this results in approaching the radiation-free direction.

Suppose that a particle has velocity $\textbf{v}$, different from the velocity $\textbf{v}_{\rm RFD}$ of the radiation-free motion by a small deviation $\Delta \textbf{v}$, such that $\textbf{v} = \textbf{v}_{\rm RFD} + \Delta \textbf{v}$. Although the effects of radiation reaction can be discrete and stochastic, for our analysis it is sufficient to consider a {continuous} effect (we stress that our general result is valid independent of this choice of the radiation reaction treatment), where the radiation reaction force is $\textbf{F}_{\rm RR} = -0.37 m^2 c \hbar^{-2}\chi^{2/3} \textbf{v}$. This gives the averaged force in the limit of $\chi \gg 1$, where $\chi = \gamma F_\perp/E_{\rm S}$ is the quantum efficiency parameter for a particle experiencing a force of magnitude $F_\perp$ across its motion, and $E_{\rm S} = m^2c^3/\hbar$ is the {Schwinger-Sauter} field \cite{DiPiazza-review}. We also note that in the relativistic regime this is the dominant part of the radiation reaction force, while the general expression contains components perpendicular to the propagation velocity. The equation for the evolution of the particles momentum $\textbf{p}$ is \\[-5mm]
\begin{equation}
  \begin{array}{l}
	\displaystyle{\dot{\textbf{p}}= \textbf{E} + \frac{1}{c}\left(\textbf{v}_{\rm RFD} + \Delta \textbf{v} \right) \times \textbf{B} + \textbf{F}_{\rm RR}  }\\
	\displaystyle{= \textbf{e}_\parallel + \textbf{F}_{\rm RR} + \left(\textbf{e}_\perp + \frac{1}{c}\textbf{v}_{\rm RFD}\times \textbf{B}\right) + \frac{1}{c}\Delta \textbf{v}\times \textbf{b}_\parallel + \frac{1}{c}\Delta \textbf{v}\times \textbf{b}_\perp,}
	\end{array}
	\label{p1}
\end{equation}
where $\textbf{e}_\parallel$, $\textbf{e}_\perp$, $\textbf{b}_\parallel$ and $\textbf{b}_\perp$ are the parallel and perpendicular components of $\textbf{E}$ and $\textbf{B}$ relative to the radiation-free direction, and the dot stands for time derivative. The expression in the brackets on the second line of Eq.~(\ref{p1}) is identically zero by the definition of $\textbf{v}_{\rm RFD}$. The next term is perpendicular to $\textbf{v}_{\rm RFD}$ and thus does not cause either approach to or deflection from this direction. The last term is almost parallel to the direction of motion, because in the case of highly relativistic motion (which we are considering here) $\Delta \textbf{v}$ is close to perpendicular to $\textbf{v}_{\rm RFD}$ and $\textbf{v}$. The first term $\textbf{e}_\parallel$ is also almost parallel to $\textbf{v}$. However, if the electric field is not zero, the last term gives a second order effect due to $\Delta \textbf{v}$ being small. Thus approaching to or deviating from the radiation-free direction is governed by the electric field component parallel to RFD, i.e. $\textbf{e}_\parallel$.

Out of two counter-orientated direction that yields zero lateral acceleration, the RFD direction (\ref{eq:rdf}) is defined so that $\textbf{e}_\parallel$ points towards the radiation-free motion, making $\Delta \textbf{v}$ smaller (with exceptions of $(E = 0)$ and $\left(E = B, \textbf{E} \bot \textbf{B} \right)$). Thus the force of $\textbf{e}_\parallel$ acts so that a particle approaches the radiation-free direction. In order to obtain the time-scale of this {behaviour} we rewrite the expression (\ref{p1}) in terms of the relativistic $\gamma$ factor and the angle $\delta$ of deviation from RFD. We obtain
\begin{equation}
	{mc \dot{\gamma} = e_\parallel \cos\delta - A \left(\gamma e_\parallel \sin\delta \right)^{2/3}} \text{ and }\, 
	{mc\gamma \dot{\delta} = - e_\parallel \sin\delta,}
	\label{p2}
\end{equation}
where $e_\parallel = |\textbf{e}_\parallel|$ and $A = 0.37 \alpha E_{\rm S}^{1/3}$, $\alpha \approx 1/137$ is the fine-structure constant hereafter. This is an autonomous system of differential equations, which is accessible for analytical study via its phase plane. However, the crucial features and estimates can be obtained with the following simple analysis. First, we assume that $\delta \ll 1$. From the first equation of (\ref{p2}) we see that the particle can gain energy until the Lorentz force (the first term) becomes balanced by the radiation reaction (the second term). The smaller {the} deviation, the higher {the} values of $\gamma$ {can be achieved}. The limiting case of balance between these terms corresponds to the radiation-dominated regime \cite{PPR,Bulanov:2010gb}. For this limit we can determine the relation $\gamma^{-1} \approx \delta \, A^{3/2} e_\parallel^{1/2}$. Using the second equation in (\ref{p2}), we obtain
\begin{equation}
\dot{\delta} = -\delta^2/\tau_{\rm RFD},
\label{d1}
\end{equation}
where $\tau_{\rm RFD}$, the typical time-scale of approaching to the radiation-free direction, is given by
\begin{equation}
\tau_{\rm RFD} = (0.37 \alpha)^{-3/2}{mc}/(e_\parallel E_{\rm S})^{1/2} \approx 7100 {mc}/(e_\parallel E_{\rm S})^{1/2}. 
\label{tau_rf}
\end{equation}
As one can see from Eq. (\ref{d1}) the particle indeed approaches the radiation-free direction ($\delta = 0$), with temporal evolution 
\begin{equation}
\label{time_scale}
\delta(t) \sim \tau_{\rm RFD}/t.
\end{equation} 

During the particle motion, the RFD evolves according to the evolution of the electric and magnetic fields experienced by the particle. Assuming that the RFD rotates with a typical frequency $\omega$ of the electromagnetic field we can determine the typical value of the deviation angle as a function of field amplitude $a$ given in relativistic units ($mc\omega/e$, where $e$ is the charge of positron):
\begin{equation}
\label{deviation_angle}
\left\langle \delta\right\rangle \sim {7100}/{\sqrt{a a_{\rm S}}},
\end{equation}
where $a_{\rm S}$ is the Schwinger-Sauter field given in relativistic units relative to the frequency $\omega$.

\textit{Insight into the origins of trapping phenomena.\textemdash}One indicative consequence of {charged particles approaching} radiation-free motion is the phenomenon of anomalous radiative trapping (ART) \cite{gonoskov.prl.2014}. The explanation given in Ref.~\onlinecite{gonoskov.prl.2014} is based on the emergence of radiation-free motion in case of a linearly polarized electromagnetic standing wave (see fig.~1~(c) in Ref.~\onlinecite{gonoskov.prl.2014}), \Arkady{and we here can estimate the threshold $a^{\rm ART}_{\rm th}$ for the appearance of this effect using Eq. (\ref{tau_rf}).} In order for particles to migrate from the vicinity of an electric field node [where the particles are accumulated by normal radiation trapping (NRT)], the radiation-dominated motion should appear in a small (enough) neighborhood around this point. According to fig.~1~(c) in Ref.\ \onlinecite{gonoskov.prl.2014} the typical spread of particles around the magnetic field node in the NRT regime is about $1/10$ of the wavelength and the typical electric field strength at this point is about 1/10 of the standing-wave amplitude. The dominance of radiation-free motion can be associated with the requirement of $\tau_{\rm RFD}$ being less than one eighth of the wave period. Expression (\ref{tau_rf}) then yields the threshold amplitude (in relativistic units)
\begin{equation}
a^{\rm ART}_{\rm th} \approx 2000 
\end{equation}
which is consistent with the threshold determined numerically in Ref.~\onlinecite{gonoskov.prl.2014} {(see Fig.\ fig.~1~(a) of that reference)}.

Another phenomenon that can be explained by the particles approaching radiation-free motion is the phenomenon of radiation reaction trapping \cite{pukhov.prl.2014}. {This} effect appears as the tendency of particles to co-propagate with an intense laser pulse. From the analysis presented here, {we see that} the role of the radiation reaction is to reduce the gamma factor of the electrons so that the {Lorentz force} can quickly deflect the particles towards \Arkady{the RFD, which in this case coincides with the wave vector.} Once the particles come close to this direction, they can propagate for a long time together with the pulse.

\textit{Radiation-dominated particle and plasma dynamics.\textemdash}For sufficiently strong fields the approach to the RFD (\ref{time_scale}) can be faster than the evolution of the field, and thus of the $\textbf{n}_\text{RFD}$, in time and along the trajectory. In this case, particles will constantly follow the local RFD at each point of space, creating currents that in turn affect electromagnetic fields. Using this as a key assumption, we can obtain a self-consistent description of such radiation-dominated plasma dynamics in the form
\begin{eqnarray}
	&& {{\partial_t f_i}+ c \textbf{n}_{\rm RFD}^\pm(\textbf{E}, \textbf{B}) \cdot \nabla f_i = 0,}\\
	&& {{\partial_t f_k} + \textbf{v}_k \cdot \nabla f_k + q_k\left(\textbf{E} + \frac{\textbf{v}_k}{c} \times \textbf{B}\right) \cdot \frac{\partial f_k}{\partial \textbf{p}_k} = 0,}
\end{eqnarray}
with $\rho = \sum_k q_k \int f_k d^3p + \sum_i q_i f_i $ and $ \textbf{j} = \sum_k q_k \int f_k \textbf{v}_k d^3p + \sum_i q_i f_i c \textbf{n}_{\rm RFD}^\pm(\textbf{E} , \textbf{B}) $  
being the charge density and current density, respectively, and the particle velocity given by $\textbf{v}_k = \textbf{p}_k\left[1 + \textbf{p}_k^2/(m_k c)^2\right]^{-1/2}/m_k$. 
{Here the index $i$ denotes particles that undergo radiation-free motion and $k$ denotes other particles (for example heavier ions), $f_i$ and $f_k$ are the respective distribution functions, and the self-consistent fields are governed by Maxwell's equations with the currents and charge densities generated by the above distribution functions.} Note that this description does not imply the absence of radiation, which can be obtained through accounting for the evolution of RFD and the gamma factor across the obtained trajectories. This description can also be extended by accounting for the related particles redistribution between the radiation-free trajectories (as a second order effect).

To analyze the capabilities of this approach and to identify its range of applicability we performed a numerical study. We considered an array of positrons moving in an arbitrary chosen configuration of the electromagnetic field that has a typical frequency $\omega$ and amplitude $a$. For this purpose, we consider eight linearly polarized plane waves each having the frequency of $\omega$ or  $\omega/2$ and propagating in positive or negative direction of $x$ or $y$ axis of a right-handed coordinate system $xyz$. \Arkady{The waves have arbitrarily chosen phases, polarization and arbitrarily chosen amplitudes with the average value of $0.475 a$}. We use dimensionless time $t$ and coordinates $x$ and $y$ that are normalized to the period $T = 2\pi/\omega$ and wavelength $\lambda = 1 \mu$m respectively; amplitude $a$ is in relativistic units.

\begin{figure}
	\centering\includegraphics[width=1.0\columnwidth]{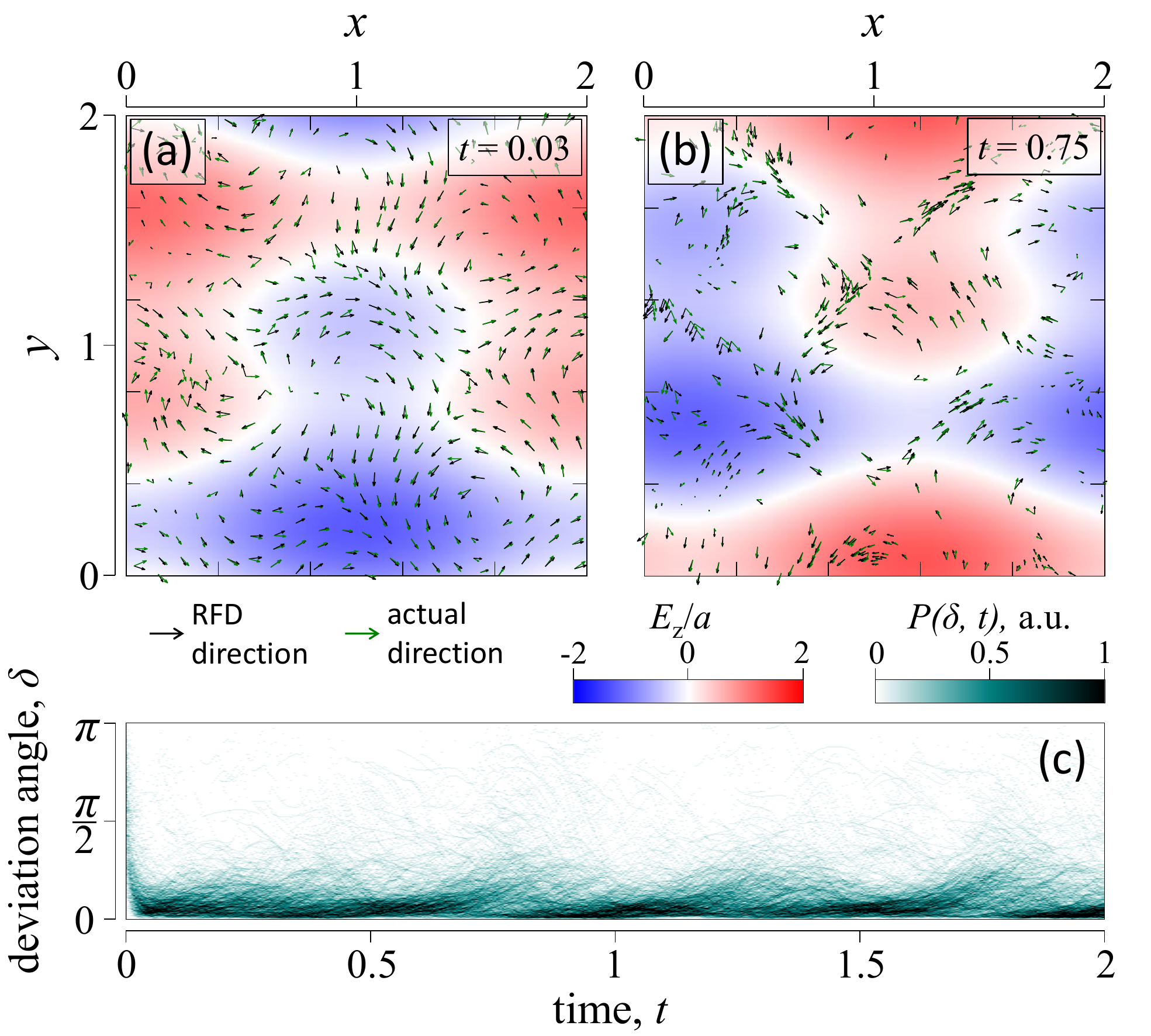}
	\caption{The result of numerical simulation of particles' motion for the case of the field amplitude $a = 10^4$. The position of particles is shown at $t = 0.03$ (a) and $t = 0.75$ (b). For each particle we show the $xy$-projection of the direction of motion (green arrow) and of the RFD calculated for the local field (black arrow). The temporal evolution of particles distribution $P(\delta, t)$ in deviation angle $\delta$ is shown in panel (c). See supplementary video for the entire evolution.} \label{rfd_example}
\end{figure}

The field structure is uniform in the $z$ direction and has a period of 2 in both $x$ and $y$ directions. Initially, the positrons have random momenta of typical scale $mca$ and are placed equidistantly within this periodic region. We employ periodic boundary conditions during their motion computed in the given field. The radiation reaction is accounted for through photon emission events that stochastically occur according to the QED rate expressions. For this purpose we use the numerical routines described in Ref. \cite{gonoskov.pre.2015}. \Arkady{We show the result of our simulation for the case of $a = 10^4$ in fig.~\ref{rfd_example}, where one can clearly see that the directions of particles motion (green arrows) systematically approach the local RFD (black arrows).}  

During the motion of a particle the local fields change smoothly in time and so does RFD most of the time. However, if $E_\parallel$ changes the sign and $E_\bot < B$ then RFD changes in a stick-slip way (see fig.~\ref{rfd_view}). \Arkady{This explains the large deviations observable for some particles in fig.~\ref{rfd_example}.} We can thus distinguish two qualitatively different regimes of motion and emission: the regime of smooth evolution of the RFD and the regime of stick-slip change of the RFD. In the former case, a particle continuously gain energy and emit it predominantly in the form of relatively low-energy photons. In the latter case, the deviation angle and the lateral acceleration instantaneously become large and this results in the emission of high-energy photons under relatively large values of $\chi$.

\begin{figure}
	\centering\includegraphics[width=1.0\columnwidth]{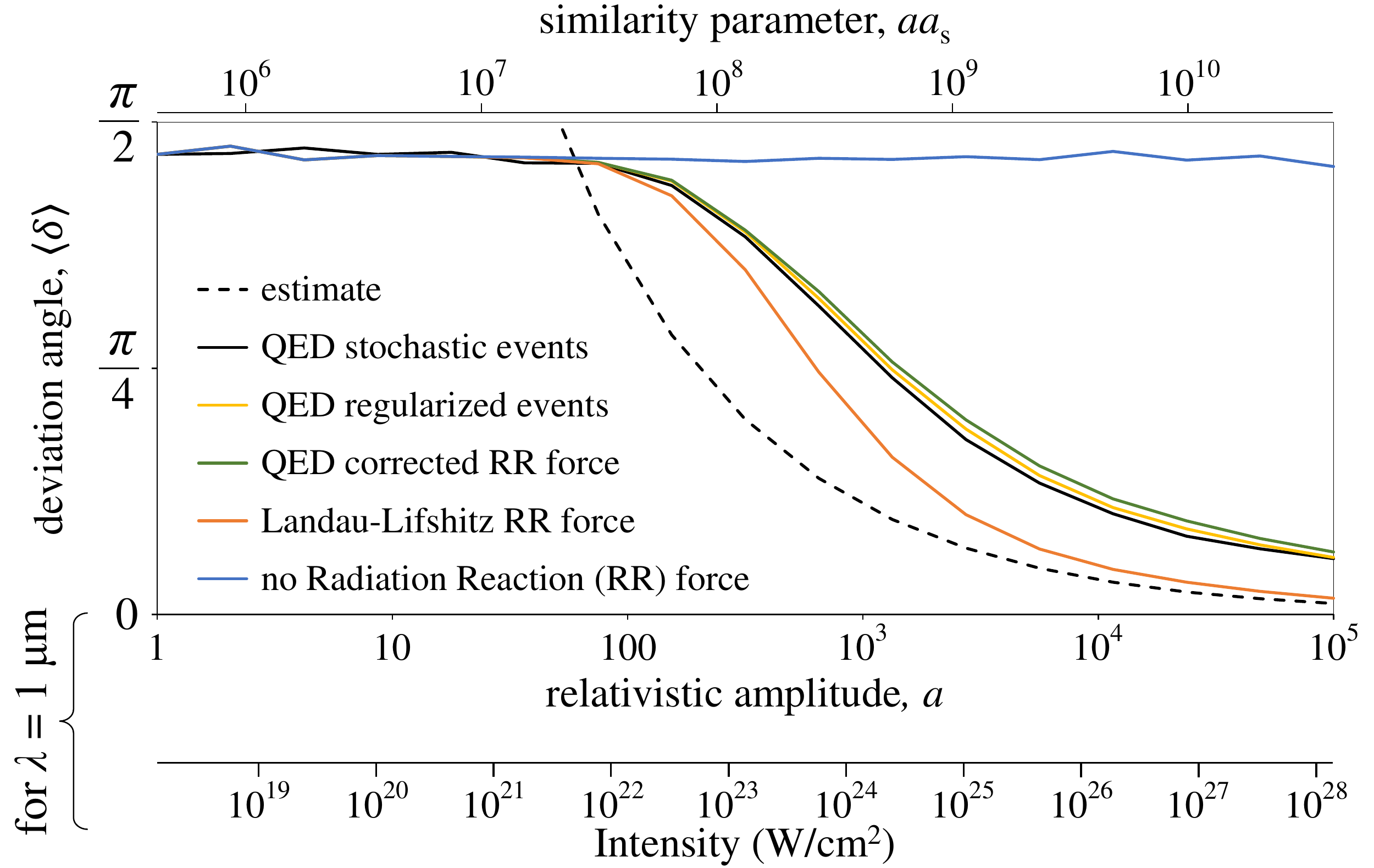}
	\caption{The dependency of the average deviation angle $\left\langle \delta\right\rangle$ on the field amplitude in relativistic units. The results obtained numerically with different models for the radiation reaction are shown with solid curves (see description on the diagram) and the result of the estimate (\ref{deviation_angle}) is shown with dashed curve.} \label{rfd_deviation}
\end{figure}

To identify the applicability of the proposed approach for different field strengths we perform the described simulations for a range of values of the amplitude $a$ and for each case we determine the average deviation angle $\left\langle \delta\right\rangle$  for the second half of the simulation. For the completeness we also apply different approaches to account for the radiation reaction. The result of this study is shown in fig.~\ref{rfd_deviation} together with the estimate (\ref{deviation_angle}). Although one can see a good agreement, the expression (\ref{deviation_angle}) systematically underestimate the typical deviation angle. One reason for this is the fact that when obtaining the expression (\ref{deviation_angle}) we do not account for the regime of stick-slip deviation from RFD. For the physics of high-intensity laser-plasma interactions (see scales for $\lambda = 1\,\mu$m) a systematic tendency of particles to approach the RFD appears at intensities on the order of $10^{23}$~W/cm$^2$. This tendency becomes overwhelming for intensities above $10^{25}$~W/cm$^2$, which are expected to be reached at the large-scale high-intensity laser facilities of the next generation.

In this paper, we have analysed particle motion in the limit of strong radiation reaction. We have shown that for arbitrary electric and magnetic field there is always one direction of propagation that leads to the absence of lateral acceleration. We also demonstrate that the particles approach {this direction within a finite, characteristic time, and we derive expressions for the representative dynamics of this approach towards the radiation-free direction.} We discuss how our conclusions can provide a possibility to analyse and understand particle and plasma dynamics in strong fields.

\textit{Acknowledgments.\textemdash}We thank Tom Blackburn and Anton Ilderton for helpful and stimulating discussions. The research is partly supported by the Russian Science Foundation Project No. 16-12-10486 (A.G.), by the Swedish Research Council grants No. 2013-4248, 2016-03329 (M.M.) and 2017-05148 (A.G.), and by the Knut \& Alice Wallenberg Foundations (M.M. and A.G.).

\end{document}